\documentclass[useAMS,usenatbib,usegraphicx]{mn2e}

\newcommand{\dm}{\mathrm {dm}}
\newcommand{\br}{\mathrm {b}}
\newcommand{\Lya}{Ly$\alpha$~}
\newcommand{\twcm}{$21$\mathrm{cm}}
\newcommand{\bk}{{\bf k}}

\title{Growth of Linear Perturbations before the 
Era of the First Galaxies}

\author[S. Naoz and R. Barkana]{S. Naoz and R. Barkana$^{1}$
\thanks{E-mail: smadar@wise.tau.ac.il (SN); barkana@wise.tau.ac.il 
(RB)}\\ $^{1}$School of Physics and Astronomy, The Raymond and Beverly
Sackler Faculty of Exact Sciences, Tel Aviv University, Tel Aviv
69978, ISRAEL}

\begin{document}

\pagerange{\pageref{firstpage}--\pageref{lastpage}} \pubyear{2005}

\maketitle

\label{firstpage}

\begin{abstract}

We calculate the evolution of linear density and temperature
perturbations in a universe with dark matter, baryons, and radiation,
from cosmic recombination until the epoch of the first galaxies. In
addition to gravity, the perturbations are affected by electron
scattering with the radiation, by radiation pressure, and by gas
pressure. We include the effect of spatial fluctuations in the
baryonic sound speed and show that they induce a $\ga 10\%$ change in
the baryonic density power spectrum on small scales, and a larger
change on all scales in the power spectrum of gas temperature
fluctuations. A precise calculation of the growth of linear
perturbations is essential since they provide the initial conditions
for the formation of galaxies and they can also be probed directly via
cosmological 21cm fluctuations. We also show that in general the
thermal history of the cosmic gas can be measured from 21cm
fluctuations using a small-scale anisotropic cutoff due to the thermal
width of the 21cm line.

\end{abstract}


\section{Introduction}

Observations of the cosmic microwave background (CMB) show that the
universe at cosmic recombination (redshift $z\sim 10^3$) was
remarkably uniform apart from spatial fluctuations in the energy
density and in the gravitational potential of roughly one part in
$10^5$ \citep[e.g.,][]{bennett}. The primordial inhomogeneities in the
density distribution grew over time and eventually led to the
formation of galaxies as well as galaxy clusters and large-scale
structure.

Different physical processes contributed to the perturbation growth
\citep[e.g.,][]{Peebles, pee, Ma}. In the absence of other influences,
gravitational forces due to density perturbations imprinted by
inflation would have driven parallel perturbation growth in the dark
matter, baryons and photons. However, since the photon sound speed is
of order the speed of light, the radiation pressure produced sound
waves on a scale of order the horizon and suppressed sub-horizon
perturbations in the photon density. The baryonic pressure similarly
suppressed perturbations in the gas below the much smaller baryonic
Jeans scale. Since the formation of hydrogen at recombination had
decoupled the cosmic gas from its mechanical drag on the CMB, the
baryons subsequently began to fall into the pre-existing gravitational
potential wells of the dark matter.

Spatial fluctuations developed in the gas temperature as well as in
the gas density. Both the baryons and the dark matter were affected on
small scales by the temperature fluctuations through the gas pressure.
Compton heating due to scattering of the residual free electrons
(constituting a fraction $\sim10^{-4}$) with the CMB photons remained
effective, keeping the gas temperature fluctuations tied to the photon
temperature fluctuations, even for a time after recombination. In
prior analyses \citep[e.g.,][]{Peebles, Ma} and in the standard
CMBFAST code \citep{cmbf} a spatially uniform speed of sound was
assumed for the gas at each redshift. This assumption meant that the
gas temperature distribution at any given time was assumed to be
simply proportional to its density distribution. While this assumption
can be accurately invoked before recombination, when the gas is
thermally and mechanically coupled to the radiation, and in the
absence of heating would be accurate after recombination (i.e., for an
adiabatic gas), this assumption is in fact inaccurate during the
post-recombination era due to the continual Compton heating. A full
calculation of the temperature fluctuations is essential for obtaining
accurate results for the small-scale density power spectra of both the
baryons and (to a lesser degree) the dark matter (see also
\citet{Naoshi1} and \citet{Naoshi2}, who considered the effect of a 
varying sound speed on density perturbations). Note that whether we
discuss perturbations driven by gravity or sound waves driven by
pressure gradients, we in every case refer to $\delta p/ \delta \rho$
as the square of the sound speed of the fluid, where $\delta p$ and
$\delta \rho$ are the pressure and density perturbations,
respectively.

The primordial inhomogeneities in the cosmic gas induced variations in
the optical depth for absorption of the CMB at the redshifted 21cm
wavelength of neutral hydrogen. Therefore, the hyperfine spin flip
transition of neutral hydrogen is potentially the most promising
tracer of the cosmic gas in the era before the first galaxies. Future
observations of the redshifted $\twcm$ power spectrum as a function of
wavelength and angular direction should provide a three-dimensional
map of the distribution of neutral hydrogen \citep{H_R,madau}, where
redshift supplies the line-of-sight distance. At different redshifts,
different signal characteristics are expected, according to the
evolution of density fluctuations and the thermal history of the
cosmic gas.

\citet{hzf} explored the consequences of the varying baryonic sound 
speed for the evolution of perturbations on sub-horizon large-scale
structure scales (wavevectors $k=0.01$--40 Mpc$^{-1}$) in the redshift
range $z=20$--150, where the analysis involves only gravity; they
showed that observations of 21cm fluctuations at these redshifts can
measure five easily-separated fluctuation modes. In this paper we
extend the redshift range back to cosmic recombination, we consider
the largest scales (\S 2) where general relativity and photon
fluctuations play a role, and we also study the smallest scales (\S 3)
where gas pressure has an essential effect. This allows us to study
the evolution of temperature fluctuations on all scales, and to
produce the first accurate predictions at $k > 100$ Mpc$^{-1}$ of the
baryonic density power spectrum and (in \S 4 \& 5) of 21cm
fluctuations. We also calculate (\S 5) a general feature of the 21cm
fluctuation power spectrum, i.e., smoothing by the thermal width of
the 21cm absorption line. We give our conclusions in \S 6.

\section{Growth of Large Scale Density Perturbations}

Up to recombination the baryons and the photons were tightly coupled,
and afterwards the baryons were significantly Compton heated down to
$z \la 200$ and retained some memory of this heating at even lower
redshifts. Thus, calculations of the growth of perturbations in the
baryon density and temperature must take into consideration the
evolution of the photon density and temperature fluctuations. The
perturbations in the photons themselves \citep[e.g.,][]{Ma} are not
significantly affected by the baryon temperature since the photon
pressure dominates strongly over the baryonic pressure.

If we define the photon temperature $T_\gamma$ at each point as the
temperature averaged over all photon directions, then dimensionless
fluctuations in this temperature ($\delta_{T_\gamma}$) are related to
photon density fluctuations ($\delta_\gamma$) by
\begin{equation} \label{eq:gamma}
\delta_\gamma=4\delta_{T_\gamma}\ .
\end{equation}
We can derive an equation for the evolution of fluctuations
$\delta_{T}$ in the gas temperature $T$ from the first law of
thermodynamics [following the derivation by \citet{bf}]:
\begin{equation}
\label{ther}
 dQ = \frac{3} {2} k_B dT - k_B T d \log \rho_{\br}\ ,
\end{equation}
where $dQ$ is the heating rate per particle and $\rho_{\br}$ is the
baryon density. In the post-recombination era before the formation of
galaxies, the only external heating arises from Thomson scattering of
the remaining free electrons with CMB photons, resulting in a heating
rate per particle
\begin{equation}
\label{h_rate} \frac{dQ}{dt} = 4 \frac{\sigma_{T}\, c} {m_e } \, 
k_B (T_\gamma - T) \rho_\gamma x_e(t) \ ,
\end{equation}
where $\sigma_{T}$ is the Thomson scattering cross section,
$\rho_\gamma$ is the photon energy density, and $x_e(t)$ is the
electron fraction out of the total number density of gas particles at
time $t$. After cosmic recombination, $x_e(t)$ changes due to the slow
recombination rate of the residual ions:
\begin{equation}
\label{chi} \frac{dx_e}{dt}=-\alpha_B(T)x_e^2(1+y)\bar{n}_H,
\end{equation}
where $\bar{n}_H$ is the mean number density of hydrogen at time $t$,
$y=0.079$ is the helium to hydrogen number density ratio, and
$\alpha_B(T)$ is the case-B recombination coefficient of hydrogen. We
assume that the residual electron fraction is uniform. Fluctuations in
$x_e$ are expected to be very small, since the gas in every region
starts out fully ionized and the final $x_e$ depends on the total
accumulated number of recombinations; most recombinations occur early
during cosmic recombination, and the recombination rate slows down
greatly once most of the gas has recombined, making $x_e$ insensitive
at that point to the gas density and temperature. E.g., since the
cosmic mean $x_e$ declines by only a factor of 1.3 between $z=100$ and
20, we estimate that in this redshift range, fluctuations in $x_e$ are
smaller than density fluctuations by a factor of $\sim 20$.

Combining equations~(\ref{ther}) and (\ref{h_rate}) we obtain
\begin{equation}
\frac{dT} {dt} = \frac{2} {3} T \frac{d \log \rho_{\br}} {dt} +
\frac{x_e(t)}{t_\gamma} \frac{\rho_\gamma} {\bar{\rho}_\gamma}\, 
(T_\gamma - T)\, a^{-4}\ , \end{equation}
where
\begin{equation}
\label{tgamma} t_\gamma^{-1} \equiv \frac{8} {3} \bar{\rho}_\gamma^0
\frac{\sigma_{T}\, c} {m_e} = 8.55 \times 10^{-13} {\mathrm{yr}}^{-1}
\end{equation}
and $a$ is the scale factor. Thus the evolution of the cosmic mean
gas temperature is
\begin{equation}
\label{mean} \frac{d \bar{T}} {dt} = - 2 H \bar{T} +
\frac{x_e(t)}{t_\gamma}\, (\bar{T}_\gamma - \bar{T})\, a^{-4}\ ,
\end{equation}
where $\bar{T}_\gamma=[2.725\ {\rm K}]/a$ is the mean CMB temperature,
and the first-order equation for the perturbation is
\begin{equation}
\label{gamma} \frac{d \delta_T} {d t} = \frac{2}{3} \frac{d
\delta_\br} {dt} + \frac{x_e(t)} {t_\gamma}a^{-4} \left\{
\delta_\gamma\left( \frac{\bar{T}_\gamma}{\bar{T}} -1\right)
+\frac{\bar{T}_\gamma} {\bar{T}} \left(\delta_{T_\gamma} -\delta_T
\right) \right\}\ . \label{eq:dT}
\end{equation}
The first term on the right-hand-side of each of these two equations
accounts for adiabatic expansion of the gas, and the remaining terms
capture the effect of the thermal exchange with the CMB.

We have numerically calculated the evolution of the perturbations by
modifying the CMBFAST code \citep{cmbf} according to these equations
for the temperature perturbations (along with other modifications on
small scales that are described in the next section). The initial
conditions for the temperature at $z \ga 10000$ are $\delta_T =
\delta_{T_\gamma}$ because of the tight thermal coupling between the
gas and the CMB. At lower redshifts, the mechanical coupling weakens
but the thermal coupling remains strong, i.e., the coefficient of the
coupling term in equation~(\ref{gamma}) remains very large, and it is
numerically highly inefficient to solve this equation directly. We
instead develop a simple approximation that may be used during this
era of thermal tight coupling. First we write the expression for $d
(\delta_T-\delta_{T_\gamma}) /dt$ using equation~(\ref{gamma}) along
with the expression for $d \delta_{T_\gamma} /dt$ as computed by
CMBFAST. As long as the thermal coupling is effective, the difference
$\delta_T-\delta_{T_\gamma}$ is very small, i.e., $|\delta_T-
\delta_{T_\gamma}| \ll |\delta_\br|$, so that the various terms in the
expression for $d(\delta_T-\delta_{T_\gamma}) /dt$ must cancel each
other nearly completely. Since this expression depends on $\delta_T$
[see the right-hand side of equation~(\ref{gamma})], we simply set
$\delta_T$ to the value that yields $d(\delta_T- \delta_{T_\gamma})
/dt \equiv 0$. We have checked that using this approximation down to
$z=900$ affects the power spectra by a fraction of a percent at most.

\begin{figure}
\includegraphics[width=84mm]{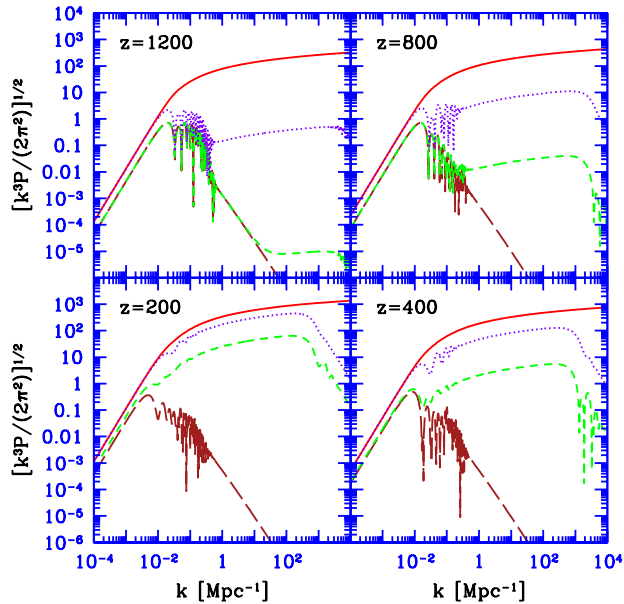}
\caption{Power spectra of density and temperature fluctuations vs.\
comoving wavenumber, at redshifts 1200, 800, 400, and 200. We consider
fluctuations in the CDM density (solid curves), baryon density (dotted
curves), baryon temperature (short-dashed curves), and photon
temperature (long-dashed curves).}
\label{fig:photons}
\end{figure}

For the concordance set of cosmological parameters \citep{s30}, with a
scale-invariant primordial power spectrum normalized to $\sigma_8=0.9$
at $z=0$, Figure~\ref{fig:photons} compares the magnitude of the
fluctuations in the CDM and baryon densities, and in the baryon and
photon temperatures. For each quantity, the plot shows the
dimensionless combination $[k^3 P(k)/(2 \pi^2)]^{1/2}$, where $P(k)$
is the corresponding power spectrum of fluctuations. Note that regions
where the fluctuations oscillate in sign (as a function of $k$) are
difficult to show precisely in such a plot (e.g., the photons and
baryons at $k=0.01$--1 Mpc$^{-1}$ at $z=1200$, and the baryon
temperature at $k>1000$ Mpc$^{-1}$ at $z=400$). Note also that the
photon temperature perturbations as shown are simply $1/4$ of the
photon density perturbations [see eq.~(\ref{eq:gamma})].

After recombination, two main forces affect the baryon density and
temperature fluctuations, namely, the thermalization with the CMB and
the gravitational force that attracts the baryons to the dark matter
potential wells. As shown in the figure, the density perturbations in
all species grow together [except that $\delta_\gamma = (4/3)
\delta_\br$] on scales where gravity is unopposed, outside the horizon
(i.e., at $k \la 0.01$ Mpc$^{-1}$ at $z \sim 1000$). At $z=1200$ the
perturbations in the baryon-photon fluid oscillate as acoustic waves
on scales of order the sound horizon ($k \sim 0.01$), while
smaller-scale perturbations in both the photons and baryons are damped
by photon diffusion (Silk damping) and the drag of the diffusing
photons on the baryons. Since the initial dark matter density
perturbations increase with $k$, while the photon perturbations are
damped on the smallest scales by photon free streaming, on
sufficiently small scales the power spectra of $\delta_\br$ and
$\delta_T$ roughly assume the shape of the dark matter fluctuation
$\delta_\dm$ (except for the gas-pressure cutoff at the smallest
scales), due to the effect of gravitational attraction on $\delta_\br$
and of the resulting adiabatic expansion on $\delta_T$.

This evolution involves two similar physical systems. In each case, a
target perturbation $\delta_0$ is driven toward one perturbation
$\delta_1$, but this forcing is opposed by coupling to a second
perturbation $\delta_2$. The values of $\delta_1$ and $\delta_2$ are
comparable on large scales but $|\delta_2| \ll |\delta_1|$ on small
scales. As long as the coupling is strong, $\delta_0 \approx \delta_2$
on large scales, while the effect of $\delta_1$ is apparent in the
form of $\delta_0$ on small scales although the coupling maintains
$|\delta_0| \ll |\delta_1|$. After the coupling weakens, the
perturbation $\delta_0$ is free to begin rising toward $\delta_1$, but
this rise occurs only gradually. In the first case,
$\delta_0=\delta_\br$ is driven by gravity toward
$\delta_1=\delta_\dm$, while mechanical coupling to
$\delta_2=\delta_\gamma$ is the opposing force. In the second case,
$\delta_0=\delta_T$ is driven by adiabatic expansion toward
$\delta_1=\frac{2}{3} \delta_b$, with resistance provided by thermal
coupling to $\delta_2=\delta_{T_\gamma}$. The mechanical coupling ends
at $z \sim 1000$ while the thermal coupling is over by $z \sim 200$.

The Figure also shows that the thermal tight-coupling approximation is
accurate at the highest redshifts shown, on large scales since
$|\delta_T-\delta_{T_\gamma}| \ll |\delta_{T_\gamma}| < |\delta_\br|$,
while on small scales $\delta_T$ and $\delta_{T_\gamma}$ are not
strongly coupled but each is individually very small compared to
$\delta_\br$. Even at somewhat lower redshifts, $\delta_\br \ll
\delta_\dm$ and $\delta_T \ll \delta_b$ on sub-horizon scales. By 
$z=200$ the baryon infall into the dark matter potentials is well
advanced and adiabatic expansion is becoming increasingly important in
setting the baryon temperature. By this redshift, the photon
perturbations are already negligible at $k \ga 0.01$ Mpc$^{-1}$,
justifying their neglect by \citet{bf} on these scales.

\section{Growth of Small Scale Density Perturbations}

On small scales (i.e., at large wavenumbers) the baryon perturbation
growth is affected by the pressure of the gas, which affects the dark
matter as well since the baryons contribute a small but significant
fraction of the total gravitational force. The evolution of
sub-horizon linear perturbations is described by two coupled
second-order differential equations. The dark matter feels the
combined gravity of itself and the baryons:
\begin{equation}
 \ddot{\delta}_{\dm}
 + 2H \dot {\delta}_{\dm} = \frac{3}{2}H_0^2\frac{\Omega_m}{a^3}
\left(f_{\br} \delta_{\br} + f_{\dm} \delta_{\dm}\right)\ ,\label{eq:dm} 
\end{equation}
where $\Omega_m$ is the redshift zero matter density as a fraction of
the critical density. The baryons feel both gravity and
pressure. Prior analyses assumed a spatially uniform baryonic sound
speed $c_s(t)$ \citep[e.g.,][]{Ma}, yielding
\begin{equation} \label{eq:cs}
\ddot{\delta}_{\br}+ 2H \dot {\delta}_{\br}=
\frac{3}{2}H_0^2\frac{\Omega_m}{a^3} \left(f_{\br} \delta_{\br} +
f_{\dm} \delta_{\dm}\right)-\frac{k^2}{a^2} c_s^2\delta_{\br}\ ,
\end{equation}
where $c_s^2=dp/d\rho$ was calculated from the thermal evolution
of a uniform gas undergoing Hubble expansion:
\begin{equation}
c_s^2 \equiv \frac{k_B \bar{T}}{\mu} \left(1 - \frac{1}{3} \frac{d
\log \bar{T}} {d \log a} \right)\ ,
\end{equation}
where $\mu$ is the mean molecular weight. This also meant that the gas
temperature fluctuation was assumed to be proportional throughout
space to the density fluctuation, so that
\begin{equation}
\frac{\delta_T}{\delta_{\br}}=\frac{\bar{c_s}^2}{k_B \bar{T}/\mu}-1\ .
\end{equation}
In this paper we instead use the equation of state of an ideal gas
to derive a more general equation for the baryons,
\begin{equation} \label{eq:b}
\ddot{\delta}_{\br}+ 2H \dot {\delta}_{\br} = 
\frac{3}{2}H_0^2\frac{\Omega_m}{a^3} \left(f_{\br} \delta_{\br} +
f_{\dm} \delta_{\dm}\right)-\frac{k^2}{a^2}\frac{k_B\bar{T}}{\mu}
\left(\delta_{\br}+\delta_{T}\right)\ .
\end{equation}
%\end{subequations}
Note that during recombination, we make a similar correction of the
baryonic pressure force in the equations of CMBFAST. In order to solve
for the density perturbations, an evolution equation for the
fluctuations in the temperature is also required. On sub-horizon
scales and at $z \la 200$ we can neglect the perturbations in the
temperature and density of the photons in equation~(\ref{gamma}),
yielding simply
\begin{equation}
\label{delta_T} \frac{d \delta_T} {d t} = \frac{2}{3} \frac{d \delta_\br}
{dt} - \frac{x_e(t)} {t_\gamma}a^{-4} \frac{\bar{T}_\gamma} {\bar{T}}
\delta_T\ . \label{eq:indian}
\end{equation}
Equations~(\ref{chi}), (\ref{mean}), (\ref{eq:dm}), (\ref{eq:b}) and
(\ref{delta_T}) are a closed set of equations describing the evolution
of density and temperature perturbations. We note that \citet{Indian}
derived a similar equation to Eq.(\ref{eq:indian}) but solved it only
for the case of a density perturbation that follows the Einstein-de
Sitter growing mode $\delta_\br \propto a$ and thus neglected spatial
variations in the speed of sound.

\begin{figure}
\includegraphics[width=84mm]{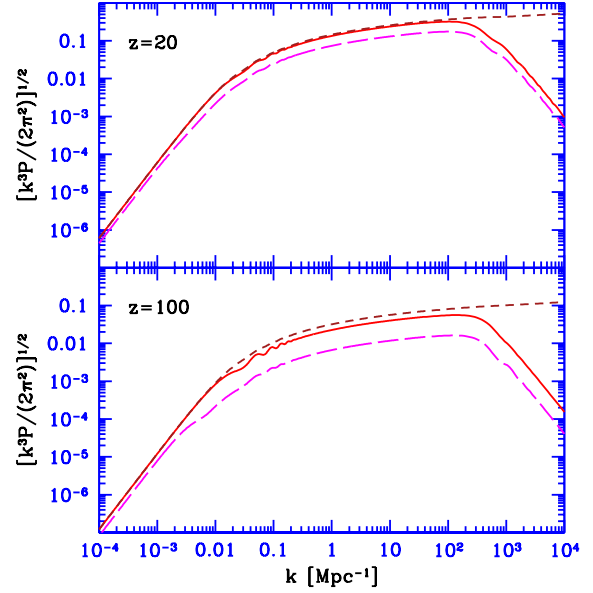}
\caption{Power spectra of density and temperature fluctuations vs.\
comoving wavenumber, at redshifts 100 and 20. We consider
fluctuations in the CDM density (short-dashed curves), baryon density
(solid curves), and baryon temperature (long-dashed curves).}
\label{fig:p}
\end{figure}

\begin{figure}
\includegraphics[width=84mm]{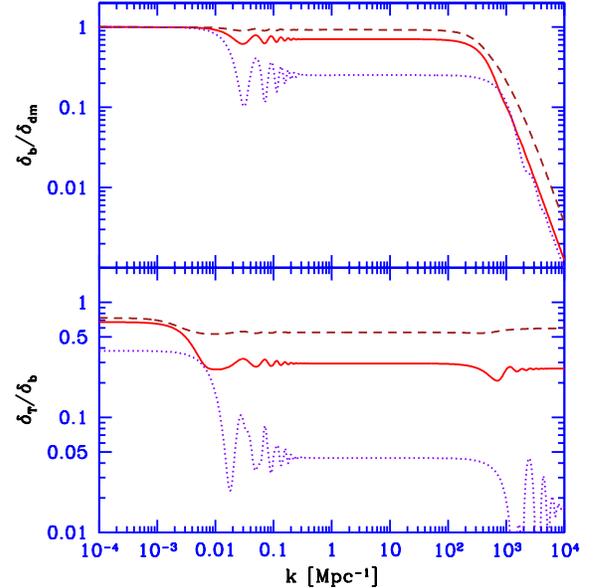}
\caption{Perturbation ratios $\delta_\br / \delta_\dm$ and $\delta_T /
\delta_\br$ vs.\ comoving wavenumber. We consider $z=400$ (dotted
curves), $z=100$ (solid curves), and $z=20$ (dashed curves).}
\label{fig:f_z}
\end{figure}

Figure~\ref{fig:p} shows the power spectra at redshift $20$ and
$100$. As time passes, the power spectrum of the baryons approaches
that of the dark matter except for the pressure cutoff, and the baryon
temperature fluctuations increase as well. However, even during the
era of the formation of the first galaxies ($z \sim 40$--20), there is
still significant memory in the perturbations of their earlier
coupling to the CMB. This is highlighted in Figure~\ref{fig:f_z},
which shows the ratios $\delta_\br / \delta_\dm$ and $\delta_T /
\delta_\br$. In both quantities, the strong oscillations that are
apparent at $z=400$ are slowly smoothed out toward lower redshifts. At
the largest scales, the baryons follow the dark matter density, and
$\delta_T / \delta_\br$ evolves from $1/3$ (the value during tight
thermal coupling to the CMB) to $\sim 2/3$ (from adiabatic
expansion). On smaller scales, the two ratios start from values $\ll
1$ during mechanical/thermal coupling, and increase towards
$\delta_\br / \delta_\dm = 1$ and $\delta_T / \delta_\br = 2/3$,
respectively. The former ratio approaches its asymptotic value
earlier, since the baryons decouple from the photons first
mechanically and only later thermally. At the smallest scales (below
the baryonic Jeans scale), the baryon fluctuation is suppressed at all
redshifts due to gas pressure, when the $k^2$ term in
equation~(\ref{eq:b}) dominates. It is clear from this figure that the
traditional assumption of $\delta_T/\delta_\br$ being independent of
scale is inaccurate at all redshifts considered here.

\begin{figure}
\includegraphics[width=84mm]{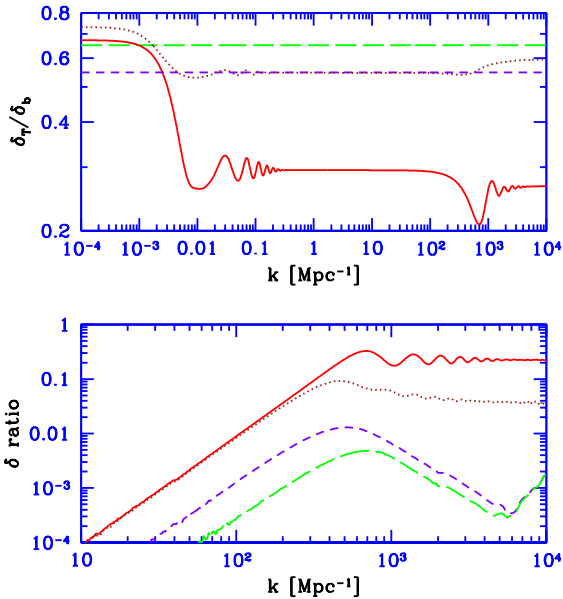}
\caption{Perturbation ratios vs.\ comoving wavenumber.
The upper panel shows the ratio $\delta_T / \delta_\br$. We consider
the improved calculation of equation~(\ref{eq:b}) at $z=100$ (solid
curve) and $z=20$ (dotted curve), compared to the traditional
calculation of equation~(\ref{eq:cs}) at $z=100$ (short-dashed curve)
and $z=20$ (long-dashed curve). The lower panel shows the ratio
between the perturbations in the improved calculation and those in the
traditional (mean $c_s$) calculation. We show the ratio of
$\delta_\br$ values at $z=100$ (solid curve) and $z=20$ (dotted
curve). Also shown is the ratio of $\delta_\dm$ values at $z=100$
(short-dashed curve) and $z=20$ (long-dashed curve). Note that the
displayed $\delta_\dm$ ratios have been smoothed at $k \ga $ 3000
Mpc$^{-1}$ to suppress noise due to limited numerical precision in
CMBFAST.} \label{fig:comp}
\end{figure}

Figure~\ref{fig:comp} shows a detailed comparison between the
fluctuation growth described by our improved equation~(\ref{eq:b}) and
that given by the traditional equation~(\ref{eq:cs}). In the improved
calculation, the ratio $\delta_T / \delta_\br$ shows different
behaviors on the scale of the horizon, the photon acoustic
oscillations, and on smaller scales. Although the ratio is roughly
constant over some ranges of scales, its value differs from that of
the traditional calculation. The improved calculation changes the
temperature fluctuations on all scales by $\ga 10\%$ at $z=20$, and by
much more at $z=100$. Note that at the lower redshift, $\delta_T /
\delta_\br$ at the small-$k$ end is higher than the adiabatic value of
$2/3$, since when $\bar{T}$ falls significantly below $\bar{T}_\gamma$
the thermal coupling tends to heat the gas more strongly in regions
with a higher photon density [see the $\delta_\gamma$ term in
eq.~(\ref{eq:dT})].

We also present in the lower panel of this Figure the ratio between
the fluctuations in the improved calculation and those in the
traditional calculation. Although the improved calculation has only a
small effect on the dark matter density ($\la 1\%$) and a similarly
small effect on the baryon density at $k < 100$ Mpc$^{-1}$, on smaller
scales the baryon fluctuations are substantially affected. The baryon
fluctuations are changed by up to $30\%$ at $z=100$ and $10\%$ at
$z=20$. Thus, accurate initial conditions for models and simulations
of the formation of the first galaxies require a full calculation of
the evolution of baryon density and temperature fluctuations along
with the dark matter.

\section{$\twcm$ Fluctuations and the Spin Temperature}

Quantitative calculations of $\twcm$ absorption begin with the spin
temperature $T_s$, defined through the ratio between the number
densities of hydrogen atoms,
\begin{equation}
\frac{n_1}{ n_0}=\frac{g_1}{g_0}e^{-T_\star/T_s},
\end{equation}
where subscripts $1$ and $0$ correspond to the excited and ground
state levels of the 21cm transition, $(g_1/g_0)=3$ is the ratio of the
spin degeneracy factors of the levels, and $T_\star=0.0682$K
corresponds to the energy difference between the levels. The 21cm spin
temperature is on the one hand radiatively coupled to the CMB
temperature, and on the other hand coupled to the kinetic gas
temperature $T$ through collisions \citep{AD} or the absorption of
\Lya photons \citep{Wout,field}. For the concordance set of
cosmological parameters \citep{s30}, the mean brightness temperature
on the sky at redshift $z$ (relative to the CMB itself) is
\citep{madau} \begin{equation} T_b = 28\, {\rm mK}\,
\left(\frac{1+z}{10}\right)^{1/2} \left(\frac{T_s - T_{\gamma}}{T_s}\right)
\bar{x}_{\rm HI},
\end{equation}
where $\bar{x}_{\rm HI}$ is the mean neutral fraction of hydrogen.

In general, fluctuations in $T_b$ can be sourced by fluctuations in
gas density, temperature, neutral fraction, radial velocity gradient,
and \Lya flux from galaxies. In this paper we consider the era before
the formation of a significant galaxy population, so that $\bar{x}_{H
I}=1$ and there is no \Lya flux. The velocity gradient term
\citep{Sobolev} is in Fourier space \citep{Kaiser,Indian2}
$\tilde{\delta}_{d_rv_r}= -\mu^2 \dot{\tilde{\delta_b } } / H $, where
$d_rv_r$ is the line-of-sight gradient of the line-of-sight gas
velocity, and $\mu = \cos\theta_k$ in terms of the angle $\theta_k$ of
$\bk$ with respect to the line of sight.  We can therefore write the
fluctuation in the brightness temperature $T_b$ as \citep{hzf}
\begin{equation}
\label{Tbk}
\tilde{\delta}_{T_b} (\bk,t) = \mu^2 \dot{\tilde{\delta}}_b(\bk) H^{-1} +
\beta \tilde{\delta}_b(\bk) + \beta_T \tilde{\delta}_T(\bk),
\end{equation}
where we have defined time-dependent coefficients $\beta$ and
$\beta_T$ [combining the relevant explicit terms from eq.~(2) of
\citet{hzf}]. Before the first galaxies, these coefficients depended
only on the cosmic mean hydrogen density $\bar{n}_H$ and the
temperatures $\bar{T}_\gamma$ and $\bar{T}$. 

\begin{figure}
\includegraphics[width=84mm]{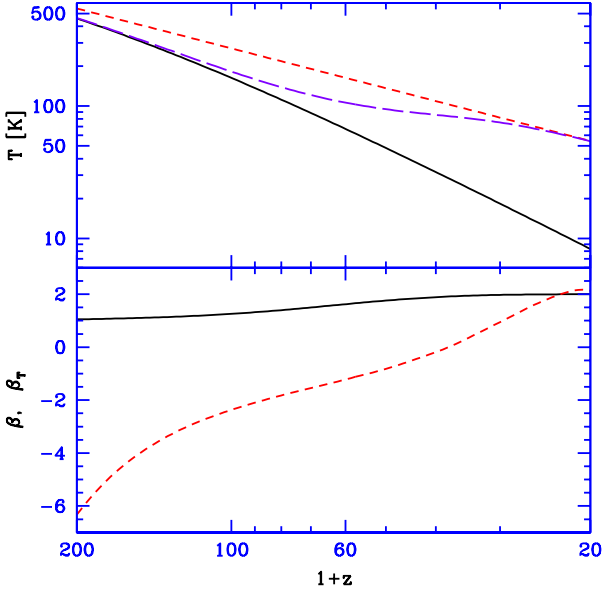}
\caption{The upper panel plots mean temperatures vs.\ redshift. We 
consider $T$ (solid curve), $T_\gamma$ (short-dashed curve), and the
$\twcm$ spin temperature $T_s$ (long-dashed curve). The lower panel
shows the $\twcm$ coefficients $\beta$ (solid curve) and $\beta_T$
(dashed curve) vs.\ redshift.} \label{fig:21cm1}
\end{figure}

Figure~\ref{fig:21cm1} shows the redshift evolution of the various
mean temperatures and the coefficients $\beta$ and $\beta_T$. As
discussed above, after recombination the small residual fraction of
free electrons coupled the temperature of the baryons to that of the
photons. Thus, after recombination the levels of the hyperfine
transition were at thermal equilibrium with the CMB temperature as
well as the temperature of the gas, i.e., $T_s\sim T\sim T_\gamma$,
and no signal is expected. At $z \la 200$ the gas temperature dropped
below the CMB temperature [$T_\gamma \propto(1+z)$], eventually
dropping adiabatically [$T\propto(1+z)^2$]. Atomic collisions in the
still-dense cosmic gas kept $T_s$ below $T_\gamma$ thus allowing the
gas to absorb at $\twcm$ against the CMB \citep{LZ}. The detection of
$\twcm$ fluctuations would uncover the thermal history of the epoch
prior to the end of cosmic reionization, and measure the growth of the
baryon density and temperature perturbations.

\section{Thermal Smoothing of the $\twcm$ Power Spectrum}

About $1\%$ of the $\twcm$ photons that propagate through the atomic
medium suffer scattering by hydrogen. Scattering at a given observed
wavelength of $21(1+z)$ cm does not occur sharply at redshift $z$, but
rather is distributed over a narrow interval about $z$. This is due to
the finite width of the $\twcm$ absorption line, set by a combination
of the natural linewidth and the thermal motions of the hydrogen
atoms. The resulting cutoff in the power spectrum was estimated
crudely by \citet{21gal}, but here we calculate it precisely along
with its angular dependence. Note that this cutoff arises from the use
of an absorption line as a probe, and is a separate effect from the
Jeans smoothing of the density discussed above.

We can obtain the observed $\twcm$ fluctuation by starting with the
previous, unsmoothed value and smoothing it along the line of
sight. The cross-section for scattering has a thermal width
$\Delta\nu_D$ given by
\begin{equation}
\Delta\nu_D=\sqrt{\frac{2k_BT}{m_H}}\frac{\nu_{21}}{c}\ ,
 \label{D}
\end{equation}
where $\nu_{21}$ is the frequency of the center of the line. The line
profile is given by the normalized Voigt profile function $\phi$,
which represents a convolution between the natural broadening and the
thermal broadening, and is given by
\begin{equation}
\phi(\xi)=\frac{u}{\pi^{3/2}}\int_{-\infty}^{\infty}
{d\eta\frac{e^{-\eta^2}}{(\xi-\eta)^2+u^2}}\ ,
\label{V}
\end{equation}
where $\xi=(\nu-\nu_{21})/\Delta\nu_D$ and
$u=A_{10}/(4\pi\Delta\nu_D)$ are the dimensionless Voigt parameters,
and $A_{10}=2.87\times 10^{-15}$ s$^{-1}$ is the Einstein spontaneous
emission coefficient for the $\twcm$ transition. The line profile
smoothes the absorption along the line-of-sight direction (comoving
coordinate $z_c$), while no such smoothing occurs in the perpendicular
directions ($x_c$ and $y_c$). It is convenient to express the result
as a smoothing with a three-dimensional window function:
\begin{equation}
\delta_{T_b}^{W}(\mathbf{r}_0 )=\int{\delta_{T_b}(\mathbf{r}- 
\mathbf{r}_0) W(\mathbf{r})d^3\mathbf{r}}\ . \label{del}
\end{equation}
In calculating the smoothing we translate frequency to comoving
coordinate $z_c$ using the homogeneous relation $(\nu_{21}/\nu)-1=a H
z_c/c$, neglecting higher-order corrections. The resulting window
function is
\begin{equation}
W(\mathbf{r})=\frac{1}{\sqrt{2} R_T} \phi(\xi(z_c)) \delta_D(x_c)
\delta_D(y_c)\ , \label{W_d}
\end{equation}
where $\delta_D$ is a one-dimensional Dirac delta function and
we have defined
\begin{equation}
R_T=\frac{1}{a H} \sqrt{\frac{k_B T}{m_H}} = 0.768
%\left(\frac{T} {\mathrm{K}} \right)^{\frac{1} {2}}
\sqrt{\frac{T} {\mathrm{K}}}
\left(\frac{\Omega_m h^2}{0.14}\right)^{-\frac{1} {2}}
\left(\frac{1+z}{10}\right)^{- \frac{1} {2}} {\rm kpc},
\end{equation}
where the second expression for $R_T$ is valid in an Einstein
de-Sitter universe (i.e., is accurate at $z \sim 3$--150). The
resulting power spectrum $P_{T_b}$ of temperature brightness
fluctuations is related to its unsmoothed value $P_{T_b}^0$ by
\begin{equation}
P_{T_b}=P_{T_b}^0 \left| \tilde{W}_k \right| ^2\ .
\end{equation}
The Fourier transform $\tilde{W}_k$ of the window function $W$ is
easily derived to be
\begin{equation}
\tilde{W}_k=e^{- \sqrt{2} k\mu u R_T} e^{-\frac{1}{2} (k\mu R_T)^2}\ ,
 \label{Wk}
\end{equation}
where $\mu=\cos\theta_k$ in terms of the angle $\theta_k$ of
$\mathbf{k}$ with respect to the line of sight. This simple
factorization of the Fourier transform is expected, since the line
profile was constructed by convolving a Lorentzian and a Gaussian, and
so the compound Fourier transform is simply the product of the
individual transforms.

The smoothing effect is characterized by two parameters, $R_T$ and
$u$. The parameter $u$ measures the relative importance of the natural
compared to the thermal line width, and its extremely low value,
$u=3.74\times10^{-19} \left(T/\mathrm{K}\right)^{-1/2}$, indicates
that the natural width is in practice negligible. We note that
collisional broadening is expected to dominate over the natural
broadening, but even collisions with other atoms, which contribute the
highest collision rate, produce broadening that is greater than the
natural width by only a few orders of magnitude, and this is still
negligible relative to the thermal broadening. Thus, the only
significant term is the Gaussian term in $\tilde{W}_k$, which cuts off
the $\twcm$ power spectrum at small scales, regardless of the physical
source of the $\twcm$ fluctuations. The cutoff drops the power
spectrum by a factor of $e$ at a wavevector $k=1/(\mu R_T)$, where,
e.g., $R_T=3.1$ kpc at $z=100$ and 1.6 kpc at $z=20$. Larger cutoff
scales are possible if the intergalactic medium is later heated by
galactic radiation; e.g., $T=10^3$ K at $z=20$ implies $R_T=17$ kpc.

Since the unsmoothed power spectrum is a polynomial in $\mu^2$
\citep{bf}, we find
\begin{equation}
P_{T_b} =
\left[\mu^4P_{\mu^4}(k)+\mu^2P_{\mu^2}(k)+P_{\mu^0}(k)\right] 
\, e^{-(k\mu R_T)^2}\ .
\end{equation}
At each $k$, the power spectrum at four different values of $\mu$
suffices to measure separately the three power spectra $P_{\mu^4}$,
$P_{\mu^2}$, and $P_{\mu^0}$, and the value of $R_T$. If many values
of $\mu$ are measured over a range of different $k$'s (including
sufficiently large $k$ values), then the thermal cutoff scale $R_T$
can be measured accurately, directly yielding a combination of $T$ and
$H$. Comparison with the three power spectra can be used to check for
self-consistency and to measure cosmological parameters, since the
power spectra also depend on $T$ [through $\beta$ and $\beta_T$ in
eq.~(\ref{Tbk})] and on the parameters (through $H$ and the growth of
the perturbations).

\begin{figure}
\includegraphics[width=84mm]{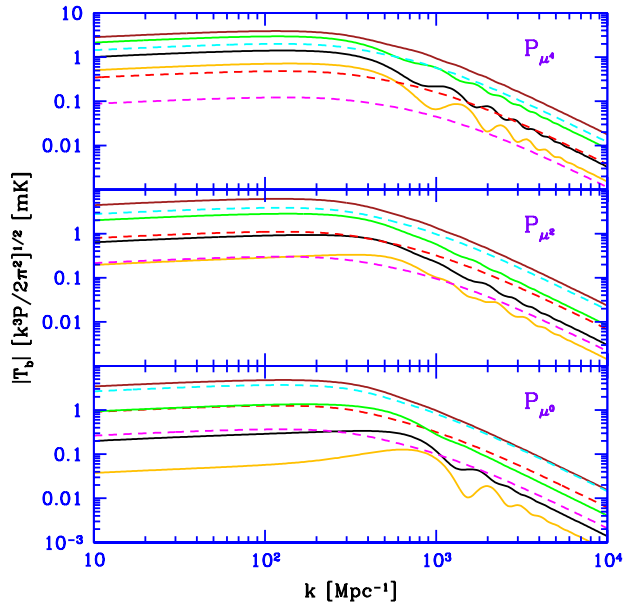}
\caption{Power spectra of 21cm brightness fluctuations versus 
comoving wavenumber. We show the three power spectra that are
separately observable, $P_{\mu^4}$ (upper panel), $P_{\mu^2}$ (middle
panel), and $P_{\mu^0}$ (lower panel). In each case we show redshifts
200, 150, 100, 50 (solid curves, from bottom to top), 35, 25, and 20
(dashed curves, from top to bottom).}
\label{fig:21cm2}
\end{figure}

\citet{hzf} showed that it would be difficult to measure 21cm
fluctuations on the largest scales ($k \la 0.01$ Mpc$^{-1}$), since
they are small and angular projections do not help. \citet{bf}
calculated the 21cm fluctuations on intermediate large-scale structure
scales, so here we focus on the predicted magnitude of small-scale
fluctuations at $k > 10$ Mpc$^{-1}$. Figure~\ref{fig:21cm2} tracks the
three 21cm power spectra from redshift 200 to 20. Each power spectrum
rises to a maximum fluctuation level of $\sim 5$ mK at $z=50$, and
subsequently drops due to the reduced collisional coupling of the 21cm
transition to the gas temperature. Small-scale oscillations are
visible at $z > 100$, with a smoother cutoff at lower redshifts. The
additional thermal smoothing (not shown in the figure) occurs at a
scale where the unsmoothed signal is $\sim 0.1$--1 mK.

\section{Conclusions}

We have computed the linear growth of fluctuations over the entire
range of scales from those outside the horizon down to the scales that
are affected by baryonic pressure. We have shown that the baryonic
sound speed varies spatially, so that the temperature and density
fluctuations must be tracked separately. At large wavenumbers ($k \ga
100$ Mpc$^{-1}$) the growth of baryon density fluctuations is changed
significantly by the inhomogeneous sound speed, by up to $30\%$ at
$z=100$ and $10\%$ at $z=20$. The effect on the dark matter evolution
is much more modest since the baryons contribute only a small fraction
of the total gravitational force felt by the dark matter.

After cosmic recombination, the gas decouples mechanically from the
photons, but remains thermally coupled down to $z \sim 200$. Starting
from very low values on sub-horizon scales, the baryon density
perturbations gradually approach those in the dark matter, and
(somewhat later) the temperature perturbations approach the value
expected for an adiabatic gas. An aftermath of cosmic recombination is
the signature of the large-scale acoustic oscillations of the
baryon-photon fluid at $k \sim 0.01$--0.2 Mpc$^{-1}$. This remnant is
most apparent in the baryon density power spectrum but it diminishes
with time. At small scales, the cutoff due to gas pressure is apparent
at all redshifts at $k \ga 10^3$ Mpc$^{-1}$, and can be measured
directly from 21cm fluctuations which reach a magnitude higher than 1
mK in the redshift range $z=100$--25.

In addition we have calculated the smoothing of the $\twcm$ power
spectrum due to the finite width of the $\twcm$ line. This effect is
dominated by the thermal motions of the scattering hydrogen atoms, and
is always present regardless of the source of the $\twcm$
fluctuations. This effect produces an anisotropic Gaussian cutoff of
the power spectrum at small scales, $k \ga 1$ kpc$^{-1}$, similar to
the characteristic wavenumber of suppression due to the baryonic
pressure. Measuring the thermal smoothing along the line of sight
requires a high resolution in frequency, e.g., $\sim 25$ Hz at $z=50$
compared to the redshifted $\twcm$ frequency of 28 MHz.

We conclude that accurate initial conditions for analytical models and
numerical simulations of galaxy formation require a full calculation
of the evolution of perturbations that includes the spatial
fluctuations in the baryonic sound speed. At high redshift the gas
temperature fluctuations are still recovering from their thermal
coupling to the photons and are quite small at sub-horizon scales;
thus, our improved calculation is particularly important for
simulations of the formation of the first galaxies.

\section*{Acknowledgments}
We acknowledge support by NSF grant AST-0204514 and Israel Science
Foundation grant 28/02/01.

%\bibitem[Gnedin \& Hui(1998)]{cs}Gnedin, N.~Y. \& Hui, L.

\bsp

\label{lastpage}

\end{document}